\documentclass[3p,times,twocolumn]{elsarticle}

\usepackage{amssymb}
\usepackage{color}
\usepackage{graphicx}
\usepackage{algorithmic}
\usepackage[plain]{algorithm}
\usepackage{listings}
\usepackage{amsmath,amssymb}
\usepackage{hyperref}

\definecolor{gray}{rgb}{0.95,0.95,0.95}
\definecolor{lightgray}{gray}{0.45}
\definecolor{myblue}{rgb}{0.22,0.22,0.61}
\definecolor{myred}{rgb}{0.6,0,0}
\definecolor{lightred}{rgb}{0.85,0.85,0.85}

\begin{document}

\begin{frontmatter}

\title{Accelerating QDP++ using GPUs}

\author[ed]{Frank Winter}

\address[ed]{School of Physics and Astronomy, University of Edinburgh,
  Edinburgh EH9 3JZ, UK}

\begin{abstract}

  Graphic Processing Units (GPUs) are getting increasingly important
  as target architectures in scientific High Performance Computing
  (HPC). NVIDIA established CUDA as a parallel computing 
  architecture controlling and making use of the compute power of
  their GPUs. CUDA provides sufficient support for C++ language elements to
  enable the Expression Template (ET) technique in the device memory
  domain.

  QDP++ is a C++ vector class library suited for quantum field theory
  which provides vector data types and expressions and forms the basis
  of the lattice QCD software suite Chroma.

  In this work accelerating QDP++ expression evaluation to a GPU was
  successfully implemented leveraging the ET technique and using
  Just-In-Time (JIT) compilation. The Portable Expression Template
  Engine (PETE) and the C API for CUDA kernel arguments were used to
  build the bridge between host and device memory domains. This
  provides the possibility to accelerate Chroma routines to a GPU
  which are typically not subject to special optimisation. As an
  application example a smearing routine was accelerated to execute on
  a GPU. A significant speed-up compared to normal CPU execution could
  be measured.

\end{abstract}


\end{frontmatter}

\section{Introduction}

GPUs are getting increasingly important in scientific HPC.
Massively multicore architectures supported by
high bandwidth memory buses make them an attractive target
architecture for either floating point operation rich or memory access
intensive applications.

NIVIDA established CUDA \cite{cuda} as their parallel computing architecture. It
enables HPC scientists to dramatically increase the computing performance
by taking advantage of the compute power of GPUs. While former releases of CUDA mostly
supported a C-like Application Programming Interface (API) with little
support for C++ features, the upcoming release 4.0 supports C++
features in a much broader way attracting an even larger set of applications and
libraries to be made subject to multicore acceleration.

Active libraries typically implemented utilising meta-programming
methods have proven to provide domain-specific abstractions to the
user and on the other hand to provide compilers with sufficient
freedom to optimise the code to a satisfactory level. 
They combine the benefits of built-in language abstractions
(convenient API, translation to efficient code) with those of 
library-level abstractions (information hiding, adaptability)
\cite{veldhuizen,pete}.

Quantum Chromodynamics (QCD) is the theory of the strong force between
gluons and quarks. The formulation of the theory on a discrete
space-time lattice is called lattice QCD and has opened the path to
numerical calculations.  Lattice QCD has received much attention as a
``grand challenge'' problem in scientific HPC. Current lattice
calculations demand computational work of sustained peta-flops and
compute resources have been and continue to be the main limiting
factor to large scale lattice QCD calculations.

Although heavily dependent on the simulation parameters typically
the largest portion of the compute time in lattice QCD
calculations is spent solving a system of linear equations when
inverting the so called fermion matrix. Typically most of the
work invested in optimisation of lattice QCD applications is spent on
this part leaving the remaining parts of the calculation fairly
unoptimised. 

GPUs form an attractive platform upon which to deploy large scale
lattice QCD calculations \cite{Clark:2009qp}. To date a lot of effort
has focussed on optimising the solver part of lattice QCD applications
\cite{Alexandru:2011ee,Cardoso:2011pw,Walk:2010ut,Chiu:2011rc}.
Outstanding implementations of several inverters which achieve a
very high sustained performance using a mixed precision approach combined with
reliable updates became available \cite{Clark:2009wm,Babich:2010mu}.
Seamless integration for these solvers are provided for a number of
lattice QCD software suites including Chroma which builds on top of
QDP++ \cite{Edwards:2004sx}.

However, as these solvers optimised for the CUDA architecture provide for
a significant speed-up of the inversion of the fermion matrix, the
remaining (unoptimised) parts of the calculation start to dominate the
overall execution time.

Currently GPU enabled software packages typically have some short
comings: A particular part of the calculation is either
executed on the GPU or the CPU leaving the respective other system
mostly idle. This is not ideal taking into account the roughly equal
acquisition cost and power consumption of both systems.

On the other hand heterogeneous computing architectures offer the
possibility to enable a cooperative computing environment where the
general purpose and specialised processors are working together in
an interleaved fashion. The idea is to split the calculation into
several small tasks and to deploy multiple types of processing
elements within a single workflow each assigned to the task its best
suited for.

In order to install this ``fine-grained'' structure in Chroma
acceleration is directly implemented in the underlying library QDP++
assigning the code parts to execute on the
processor element it is best suited for, i.e. IO-intensive operations
on the CPU, floating-point rich operations on the GPU. In this way not
only a finite set of selected functions are executed on the
accelerator but acceleration is applied in a more general fashion.

Benchmark measurements confirmed that solely accelerating the
floating-point rich operations yet yields a significant
speedup factor compared to unaccelerated execution.


The data layout, the order of access, the precision of the primitives
are left unchanged just as prescribed by the QDP++ standard template
order.

Section \ref{section:cuda} introduces briefly the CUDA architecture.
The ET technique is briefly outlined in section \ref{section:ET}.
Section \ref{section:qdp} introduces QDP++.
Section \ref{section:qdp4multicore} introduces the design elements
introduced to QDP++ for accelerating the evaluation.
Section \ref{section:API-new} details on the new QDP++ API elements.
An application example is detailed in section \ref{section:jacobi-new}.
Section \ref{section:benchmark} details on benchmarking results.

\section{The CUDA Architecture}\label{section:cuda}

The CUDA architecture is built around a set of multi-threaded Streaming
Multiprocessors (SM) which provide the main compute power.
Threads are organised in a hierarchical manner: A kernel grid is a
collection of thread blocks. A thread block is a collection of threads
and represents an indivisible unit allocatable by a multiprocessor.
Whether a given thread block can be allocated by a SM depends on the
resources (number of registers and shared memory) its threads
collectively require and the resources available on the SM. The
threads of a thread block execute concurrently on one SM, and multiple
thread blocks can execute concurrently on one SM (one block active at
a time).

The Streaming Multiprocessors implement the Single-Instruction,
Multiple-Thread (SIMT) architecture. Threads resident on one SM are
bundled into groups of 32 parallel threads, so called warps. The
threads of exactly one warp are executed in SIMT fashion. Individual
threads composing a warp start together at the same program address,
but they feature their own instruction register and are free to branch
independently. However, in this case execution is serialised.

The SM is able to hide latency to device memory by switching execution
to a different warp whose instruction is ready to execute. It is
therefore beneficial to organise the thread
number per SM in such a way that a sufficient number of warps is
resident to (ideally) completely hide the latency to device memory.

\section{Expression Templates}\label{section:ET}

C++ function and class templates together with function and operator
overloading offer the possibility to represent expressions as C++
types. This technique is commonly referred to as Expression Templates (ETs)
and was first introduced by Todd Veldhuizen \cite{exprtemplates} and
David Vandevoorde.

ETs provide a means to eliminate the need for creating temporaries when
implementing a C++ vector class library which features both a
convenient API such as for domain specific languages desired and a
high performance of the translated code. However, the latter feature
relies on the optimisation abilities of the compiler and is not
ensured in all ET applications, e.g. for Basic Linear Algebra Subprograms
(BLAS) Level 2 and 3. Here different, but also ET based approaches may
achieve a better performance \cite{2011arXiv1104.1729I}.

The Portable Expression Template Engine (PETE) pioneered the use of
the ET technique for parallel physics computations \cite{pete}.
It is an extensible implementation of the ET technique and achieves an
exceptional level of abstraction without sacrificing performance and
provides the core functionality (on the vector level) of QDP++.

Implementation of a vector library utilising the ET technique
typically involves defining a template of the evaluate function. Upon
assignment of an expression the compiler generates an 
instantiation of the evaluate function matching the expression. The
evaluate function then typically implements a loop iterating over all
vector components -- no temporary vector objects are required.

C++ compilers offer the possibility to access the function template's
arguments in a fully instantiated C++ type in form of a C string -- so
called pretty printing, which provides a means to access at runtime the
expressions.

\section{QDP++}\label{section:qdp}

QCD Data Parallel (QDP++) is a C++ vector class library suited for
quantum field theory. It forms the basis for the widely used lattice
QCD software suite Chroma and as such provides the lattice wide data
types and expressions used in Chroma \cite{Edwards:2004sx}. Chroma
implements linear algebra operations which may include nearest
neighbour communications utilising the QDP++ API.

Although not designed originally for multicore acceleration, this work
demonstrates that design elements can be added to QDP++ in such a way
that evaluations of arbitrary expressions are accelerated and
executed on a GPU. This approach was previously established when
deploying lattice QCD applications to the QPACE supercomputer
\cite{winter-phd,Nakamura:2011cd}.

\section{Accelerating QDP++ evaluation on a GPU}\label{section:qdp4multicore}

Accelerating QDP++ expressions on a GPU relies on leveraging the ET
technique to the device memory domain. A crucial component for ETs to
work in a particular memory domain is (besides the compiler's ability
to handle C++ templates) the ability to take memory addresses of
functions and allowing for dynamically
dereferencing function addresses (function pointers).

The new release 4.0 of CUDA on devices with compute capability no less
than 2.0 meets the requirements for the ET technique to work on the
device memory domain.

Unfortunately CUDA provides only a C-interface to kernel functions
making it impossible to directly pass C++ constructed expressions as
arguments to compute kernels. 

This work circumvents the aforementioned lack of a C++ API to kernel
arguments by first constructing in device memory domain an object of an
equivalent C++ expression type as used during host code translation and
second deploying the missing runtime configurable Plain Old Data (POD)
parts by copying those from the host side expression object.

In order to construct the object in device memory domain a JIT
compilation of CUDA kernel modules is triggered upon expression
evaluation. Launching this kernel constructs the required object in
device memory. Still the runtime configurable parts are missing.

To build the bridge between the two expression objects residing in
different memory domains the POD part of the expression object on host
side is copied into a C API compatible form which is allowed to be
passed as CUDA kernel arguments.

\subsection{Dynamic Code Translation, Just-In-Time Compilation}\label{section:JIT}

Entering the evaluation function triggers pretty printing of its
arguments and executing an external code generator which generates C++
device code leveraging the ET technique. The NVIDIA Compiler NVCC is
invoked which builds a shared library containing the CUDA kernel.

The shared library is loaded via the dynamic linking loader and the
kernel is executed on the device. 

After evaluation the shared library is kept loaded until the application
exits. This ensures that each kernel function is only generated once
and subsequent calls branch to the already loaded shared library.

\subsection{Flattening the Expression Tree}\label{section:flattening}

PETE \cite{pete} provides means to traverse the expression tree and
execute custom operations on the tree nodes and leaves. This method is
used on host side to collect the runtime configurable data (POD
portion) of each operator and to store them into a linear storage
container that can be passed (as a pointer) to the CUDA kernel as an
argument, i.e. the expression tree is flattened. 

The inverse operation restores the expression tree on device
side\footnote{For unknown reasons the NVIDIA Frontend++ traverses the
  expression tree in a mirrored sense compared to the GNU C++
  Compiler.}.

Certain runtime configurable operators require special
treatment. E.g. the shift operation requires read-only access to 
a site table index initialised at runtime.

A device storage container was added in such a way that the first call
to a particular runtime configurable operation triggers copying of the required
data to device memory. The storage container keeps track of already
transferred memory regions and subsequent calls to the same operation
do not trigger the device copy again but make usage of
the already resident data in device memory. The site tables remain in
device memory until they are explicitly freed by the user. This speeds
up repeated calls to the same shift function.

\subsection{Mixed Memory Domain Approach}\label{section:memory}

Since device memory is (still) a scarce resource a mixed memory
domain approach was favoured: Memory allocation for lattice wide
objects utilise the host memory domain. Upon user request the object's
data is pushed to the device memory domain.

A new feature coming with version 4.0 of CUDA provides the possibility 
to page-lock a memory range that was already allocated (4kB aligned)
in the host memory domain and to add it to the tracking mechanism to
automatically accelerate calls to device copy functions. This
mechanism eliminates the previously required staging of data regions
prior to the transfer to device memory and reduces pressure on host
memory.

\begin{lstlisting}[float,numbers=left,frame=tb,caption={Modified 
      Chroma implementation for Jacobi smearing. Prior to any calculation the 
      lattice wide objects are pushed to the device (first darker grey 
      shaded region). After calculation the result object is copied
      back to host memory and device memory is freed (second shaded
      region). QDP++ expressions (line number):
      $E_0(16,20)$, $E_1(25)$,
      $E_2(27)$,$E_3(30)$,$E_4(33)$.},label={lst:jacobi-smearing-new},escapechar=!] 
  template<typename T>
  void jacobiSmear(const multi1d<LatticeColorMatrix>& u, T& chi, 
  const Real& kappa, int iter, int no_smear_dir, const Real& _norm)
  {
    T psi;
    Real norm;
    T s_0,h_smear;

    !\colorbox{lightred}{psi.pushToDevice();}!
    !\colorbox{lightred}{for(int mu = 0; mu < Nd; ++mu )  }!
    !\colorbox{lightred}{      u[mu].pushToDevice(); }!
    !\colorbox{lightred}{chi.pushToDevice();}!
    !\colorbox{lightred}{h\_smear.pushToDevice();}!
    !\colorbox{lightred}{s\_0.pushToDevice();}!

    s_0 = chi;
      
    for(int n = 0; n < iter; ++n)
    {
      psi = chi;
      bool first = true;
      for(int mu = 0; mu < Nd; ++mu )
      {
        if (first)
        h_smear =  u[mu]*shift(psi, FORWARD, mu) + shift(adj(u[mu])*psi, BACKWARD, mu);
        else
        h_smear += u[mu]*shift(psi, FORWARD, mu) + shift(adj(u[mu])*psi, BACKWARD, mu);
        first = false;
      }
      chi = s_0 + kappa * h_smear;
    }

    chi /= _norm;

    !\colorbox{lightred}{chi.popFromDevice();}!
    !\colorbox{lightred}{for(int mu = 0; mu < Nd; ++mu ) }!
    !\colorbox{lightred}{      u[mu].freeDeviceMem();}!
    !\colorbox{lightred}{psi.freeDeviceMem();}!
    !\colorbox{lightred}{h\_smear.freeDeviceMem();}!
    !\colorbox{lightred}{s\_0.freeDeviceMem();}!

  }
\end{lstlisting}

\subsection{Thread Geometry}\label{section:thread-geometry}

The evaluation function template in ET based vector libraries
typically triggers execution of a loop iterating over all lattice
sites. With CUDA, parallelisation of a loop is typically carried out
unrolling the loop and starting one thread per loop iteration.

Since here the applied CUDA kernels not only consist of processing the lattice
sites but also require prior reconstruction of the expression tree it
was not clear that the typical approach leads to the best
performance. A software configuration parameter $N_\mathrm{site}$ was
introduced which specifies the number of sites assigned to one
thread.

CUDA kernel functions are launched with specifying the grid and block
geometries. Thus a software configuration parameter
$N_\mathrm{threads}$ 
is introduced that specifies the number of threads per block.

Given the total number of lattice sites the grid geometry is
a function of $N_\mathrm{threads}$ and $N_\mathrm{site}$.

For each expression $E_i$ a separate CUDA kernel is generated.
Thus the sustained performance $P$ is a function of 
the number of threads per block,
the number of lattice sites processed per thread, and the expression $E_i$:
\begin{equation}
P(N_\mathrm{threads},N_\mathrm{site},E_i).\nonumber
\end{equation}

CUDA enabled software packages might be equipped with an
auto-tuning mechanism that determines the optimal grid and block
geometries for the particular set of installed devices. Auto-tuning is
run prior to production and the geometry parameters that yield the
best performance are stored for later inclusion.

\section{New QDP++ API Elements}\label{section:API-new}

The QDP++ API was extended by the following elements:
\begin{itemize}
\item \texttt{OLattice::pushToLattice()} \\
Allocates a memory region of the object's size in device memory and
copies the object's data to the device.
\item \texttt{OLattice::popFromLattice()} \\
Copies the data from the device to host memory and frees device memory.
\item \texttt{OLattice::freeDeviceMem()} \\
Frees device memory.
\item \texttt{theDeviceStorage::freeAll()} \\
Frees device memory previously allocated for runtime configurable operators.
\end{itemize}

\section{Application Example: Jacobi Smearing}\label{section:jacobi-new}

Frequently used Chroma routines which are typically not subject to
special optimisation and which can consume a significant amount of
(wallclock) time when executed on a few CPU cores only are quark
smearing routines. These are operations acting on lattice
wide objects and are typically iterative prescriptions including
nearest neighbour communications. One of these routines implements
Jacobi smearing \cite{PhysRevD.47.5128} and serves here for the
benchmarking analysis. 

The Jacobi smearing procedure is obtained by solving the Klein-Gordon
equation
\begin{equation}
 K(x,x')S(x',0) = \delta_{x,0}
\end{equation}
where
\begin{equation}
 K_{x,x'}  = \delta_{x,x'} - \kappa_S \sum_{\mu} U_{\mu}(x)
\delta_{x',x+\mu}+U^{\dagger}_{\mu}(x-\mu)\delta_{x',x-\mu}
\end{equation}
as a power series in $\kappa_S$ stopping at some finite power $N_\mathrm{smear}$.

Listing.~\ref{lst:jacobi-smearing-new} shows the Chroma implementation
of Jacobi smearing using the QDP++ API . Five QDP++ expressions are involved:
$E_\mu$, with $0 \le \mu \le 4$ where the most compute intensive ones
are $E_1$ and $E_2$.

The code lines shaded darker grey were
introduced to enable execution on the GPU. All lattice objects are
pushed to the device. After calculation the device memory is freed and
the result objects copied back to its original location in host memory.

\section{Benchmark Results}\label{section:benchmark}

\begin{table}
\begin{center}
\begin{tabular}{ ccccc }
 & \multicolumn{2}{c}{SP} & \multicolumn{2}{c}{DP$^{(*)}$} \\
\hline
lattice size & CPU & GPU & CPU & GPU  \\
\hline
$8^3 \times 16$    &  1.55 & 1.64 & 1.45 & 1.49 \\
$12^3 \times 24$   &  1.52 & 6.53 & 1.40 & 4.41 \\
$16^3 \times  32$  &  1.50 & 11.86 & 1.40 & 6.26 \\
$20^3 \times  40$  &  1.52  & 16.56 & 1.41 & 7.51 \\
$24^3 \times  48$  &  1.52 & 19.09 & 1.38  & 7.91 \\
$32^3 \times  64$  &  1.51  & 21.31 & & \\
\hline
\end{tabular}
\end{center}
\caption{
  Benchmark results for the Jacobi smearing routine executed on the
  CPU and the GPU. Numbers in sustained GFLOPS for the whole smearing
  routine. $^{(*)}$Double precision throughput on this device is restricted.
  \label{table:benchmark}}
\end{table}

\begin{figure}[ht]
\begin{center}
\includegraphics[width=1.0\columnwidth]{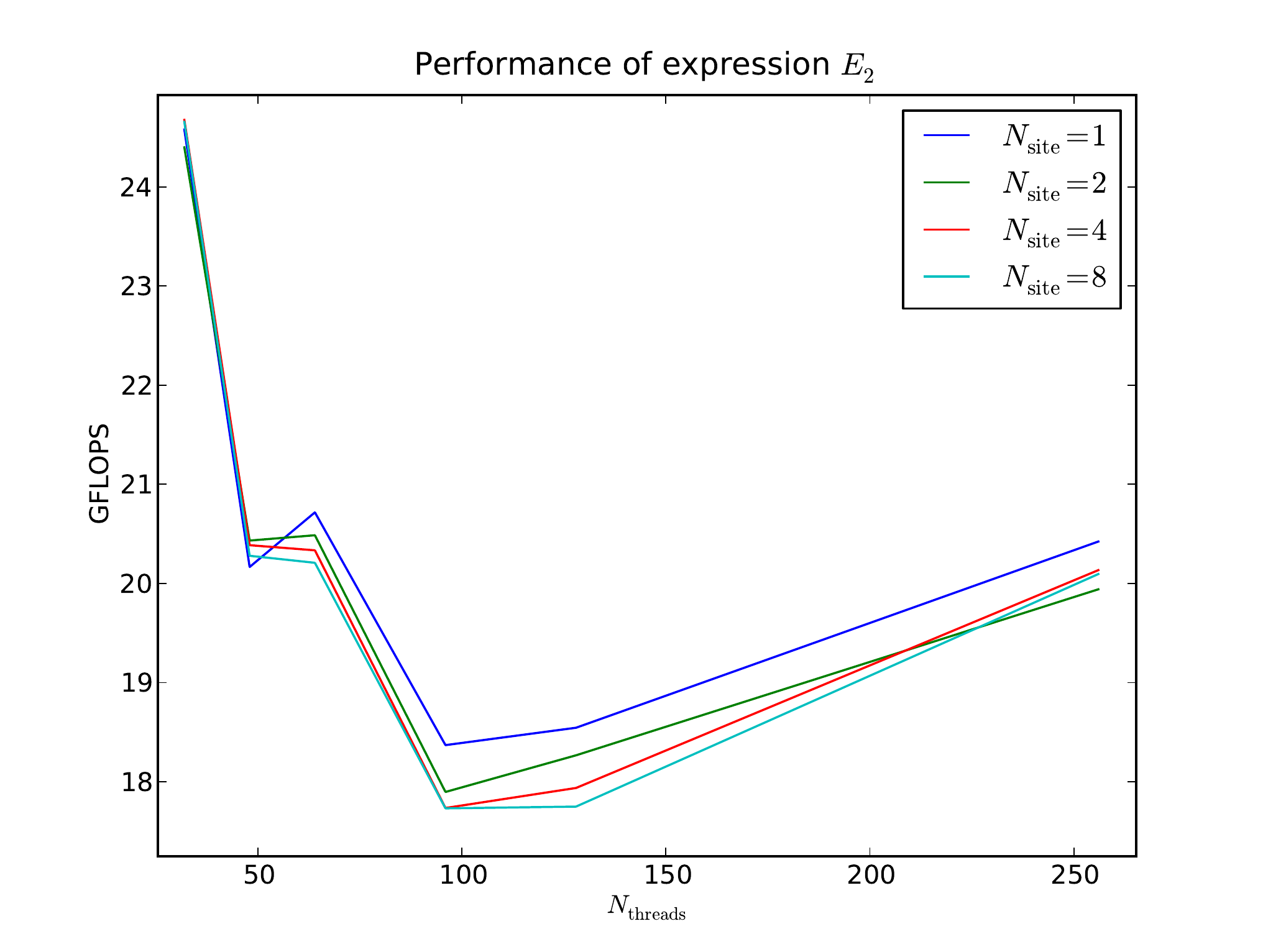}\hfill
\end{center}
\vspace*{-3mm}
\caption{\label{fig:perf-e2-nthreads} Performance dependence of
  expression $E_2$ on $N_\mathrm{threads}$ for a $L_x/a=32$ lattice
  (single precision).
}
\end{figure}

\begin{figure}[ht]
\begin{center}
\includegraphics[width=1.0\columnwidth]{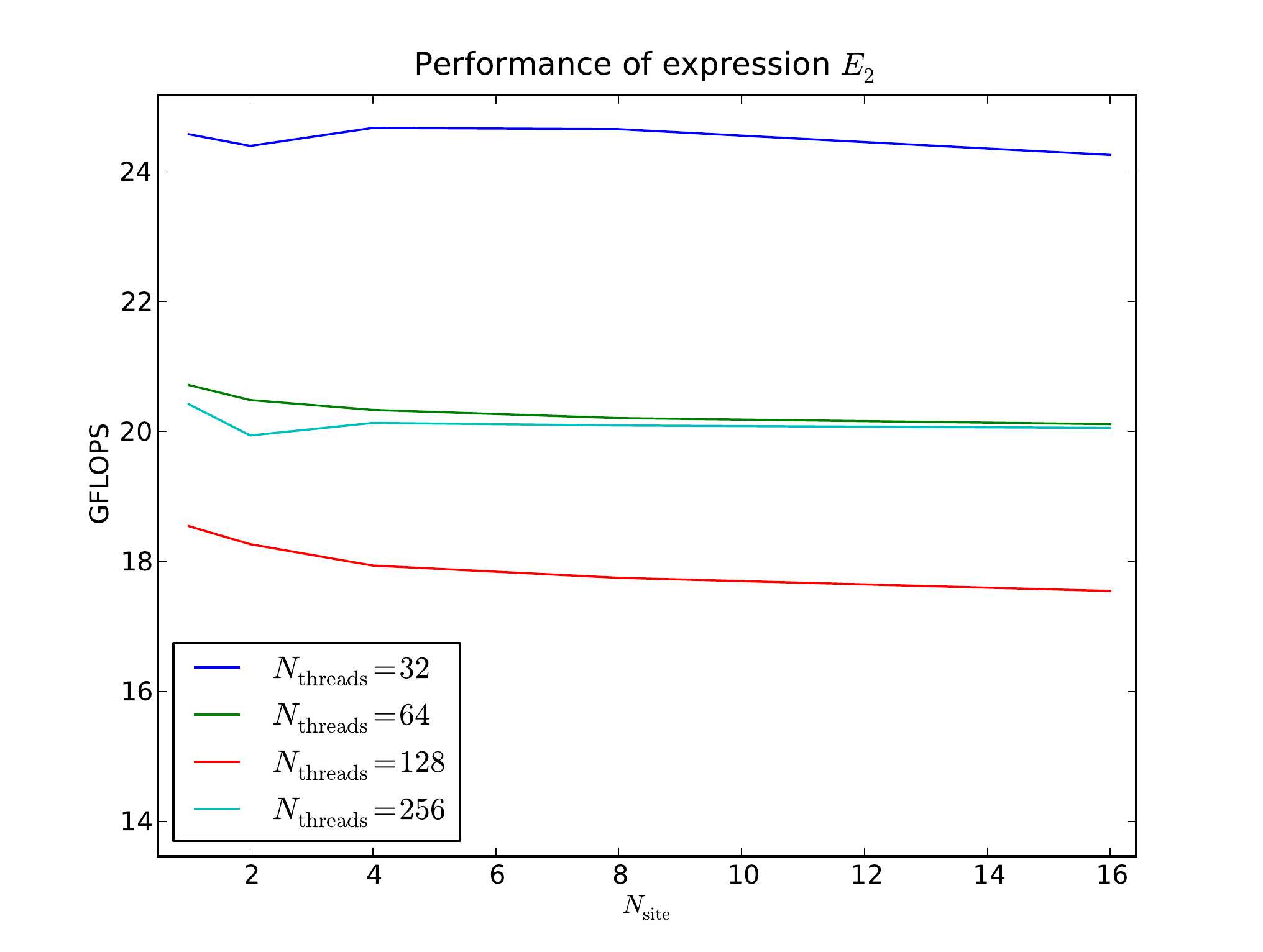}\hfill
\end{center}
\vspace*{-3mm}
\caption{\label{fig:perf-e2-nsite} Performance dependence of
  expression $E_2$ on $N_\mathrm{site}$ for a $L_x/a=32$ lattice
  (single precision).
}
\end{figure}

\begin{figure}[ht]
\begin{center}
\includegraphics[width=1.0\columnwidth]{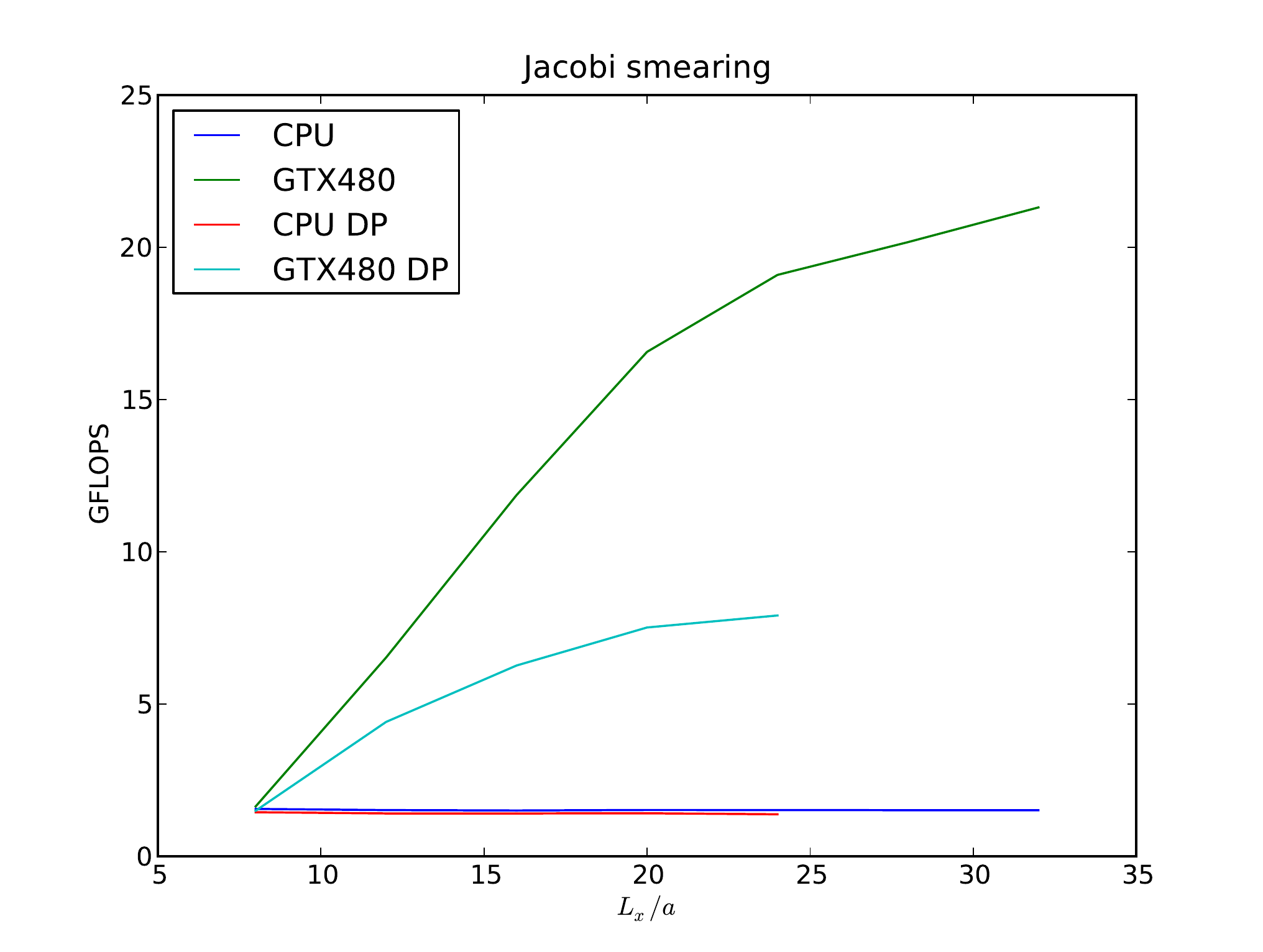}\hfill
\end{center}
\vspace*{-3mm}
\caption{\label{fig:jacobi-new}Benchmark result for Jacobi smearing on a
  NVIDIA GeForce GTX 480 in comparison to the Intel Xeon CPU. The
  number of lattice sites is given by $N = 2(L_x/a)^4$.
}
\end{figure}

Benchmarking analyses were carried out using a NVIDIA GeForce GTX 480.
This device has 1.5 GB of memory, compute capability 2.0, 15
Streaming Multiprocessors and a total of 480 CUDA cores. Double
precision on this device is restricted as it utilises the GF100 chip
designed for the consumer market. The NVIDIA CUDA 4.0 toolkit (Release
Candidate 2) was used with the NVIDIA Linux kernel driver version
270.40. 

The Chroma Jacobi smearing routine was applied to lattice objects
where lattice sizes ranged
from $N=8^3\times16$ to $N=32^3\times 64$ in single precision and
from $N=8^3\times16$ to $N=24^3\times 48$ in double precision.

The first benchmark analysis studied the performance
$P(N_\mathrm{threads},N_\mathrm{site},E)$ for varying
$N_\mathrm{threads}$ keeping $E=E_2$ and $N_\mathrm{site}=1,2,4,8$
fixed for a $32^3\times 64$ lattice in single precision. 
Fig.~\ref{fig:perf-e2-nthreads} shows the result. The best
performance was achieved when setting the thread block size equal to
the warp size $N_\mathrm{threads}=32$.

The second benchmark analysis focused on the performance dependence on 
$N_\mathrm{site}$ keeping $E=E_2$ and
$N_\mathrm{threads}=32,64,128,256$ fixed for a $32^3\times 64$ lattice
in single precision. Fig.~\ref{fig:perf-e2-nsite} shows the
result. Only a moderate dependence was seen on this configuration
parameter. Although when configuring for 32  threads a
slightly better performance is achieved when using $N_\mathrm{site}=4$
or $8$ instead of $N_\mathrm{site}=1$. 

These benchmark analyses were repeated for the expressions
$E=E_0,E_1,E_3,E_4$. 
The same characteristics were observed:
Setting the parameter $N_\mathrm{threads}=32$ resulted in
all cases clearly to the best performance with only a moderate dependence
on $N_\mathrm{site}$. 

A final benchmark analysis was carried out for the whole smearing
routine (including all five expressions) for different lattice sizes
(single and double precision) in comparison to executing the same
routine on the host CPU, an Intel Xeon CPU (E5507, 4 cores, 2.27GHz).
Tab.~\ref{table:benchmark} shows the benchmark results in numbers
(sustained GFLOPS for single and double precision).
Fig.~\ref{fig:jacobi-new} shows the result graphically.
For the $32^2\times 64$ lattice in single precision a speedup factor 
(compared to the CPU) of more than $14$ was measured.

\section{Conclusion, Outlook and Discussion}

As a first step, QDP++ expression evaluation was accelerated on the
GPU by leveraging the ET technique on the device memory domain. Solely
with acceleration a significant speedup factor for the evaluation
could be achieved.

Providing an auto-tuning mechanism for the approach described here is
not straight forward since the individual expressions $E_i$ are not
known prior to production. Establishing an auto-tuning mechanism forms
part of the to-do list. However, as the benchmark measurements showed
there seems to be a preferred thread geometry that leads to the best
performance.

Optimisation would be a next major step. By changing the data layout
in such a way that coalescing device memory access takes place an even 
higher speedup factor is expected.

A next step in another direction would be parallelisation to multiple
GPUs per host and targeting for the parallel architecture of QDP++ to
extend this approach to multiple hosts.

One might also want to move the shared linking loader to an external
daemon program or the system loader. In this way the JIT compilation
takes place only once -- even across several Chroma runs.

Utilising QDP++ profiling would eliminate the dependence on an external
code generator.

\section*{Software Availability}

QDP++ and Chroma are available as open source software
\cite{Edwards:2004sx}. QDP++ configurable for GPU evaluation is
available \cite{github}.

The GPU portion of QDP++ requires the upcoming release 4.0 of NIVIDA
CUDA \cite{cuda} and devices with compute capability lo less than
2.0 are required.

\section*{Acknowledgements}

FW is supported through a Marie Curie Early Stage Researcher
fellowship as part of STRONGnet (EU grant 238353).

\vspace{6mm}

\bibliographystyle{elsarticle-num}
\bibliography{paper}

\end{document}